# LH*TH: New fast Scalable Distributed Data Structures (SDDSs)

Dr ARIDJ Mohamed [1],

[1] Hassiba Benbouali University of Chlef Algeria

**Abstract**
Proposed in 1993 the Scalable Distributed Data Structures (SDDSs) became a profile of basis for the data management on Multi computer. In this paper we propose an organization of a LH* bucket based on the trie hashing in order to improve times of different access request.

*Keywords:* Distributed hashing, SDDS, multi computers, distributed system, access method

## 1. Introduction

A multi computer consists of set of workstations and PCs interconnected by a high speed network (Ethernet, TM Token Ring...).

It is well known that multi-computers offer best price-performance ratio; offering some new perspectives thus to high performances applications .

In order to permit the export of these performances, a new class of data structures has been proposed. It is called Scalables Distributed Data Structures (SDDS) [10] they are based on client/server architecture.

This new structure supports the parallel treatment the address computations do not involve any centralized directory. Data are typically stored in the distributed main memory (DRAM). An SDDS may easily handle many GByte files, accessible in a fraction of the disk access time. An SDDS scales to new sites through splits of those that fill up. Splits are transparently for the applications. All SDDSs support the key searches; some offer the range searches or multikey searches, Every client has his own picture of the file. The update stakes of the file structure are not sent to clients of a synchronous manner. A client can make an addressing error of then by following as result of incorrect picture.

Every server verifies the address of the received request. It is routed toward another server if an address error is detected. The adequate server sends an adjusting message to the client having made the address error, this message is called: a Picture Adjustment Message (PAM).

The PAM allows the client to adjust his picture in order not to redo the same error. This picture is not nevertheless necessarily globally exact.

Several SDDSs have been proposed. Historically, the first family is based on the hashing: DDH [6], LH* [10]. It gave rise to numerous variants, notably to high-availability [11] [13] [7] [12].

Another family has been conceived for the ordered files [1] [3],[4], [14], [6].

In this article, we present a new SDDS baptized LH*TH that consists in indexed articles of a LH* bucket, by the trie hashing [Lit 81], In order to improve the access times of different s operations.

Sections 2 and 3 of the paper recall principles of the LH* and TH respectively. The section 4 describes the principle and the organization of new SDDS LH*TH. The section 5 is dedicated to performances of the SDDS LH*TH, A comparative survey between LH*TH and LH * is presented also in this section. In we conclude this article in section 7.

## 2. The LH* SDDS

LH * [10] is a SDDS based on the linear hashing LH [8]. The extension of LH to LH * consists in putting every file bucket on a different servers of multi computer (Fig. 1). The i level of the hashing function is stocked in the headline of every LH* bucket.

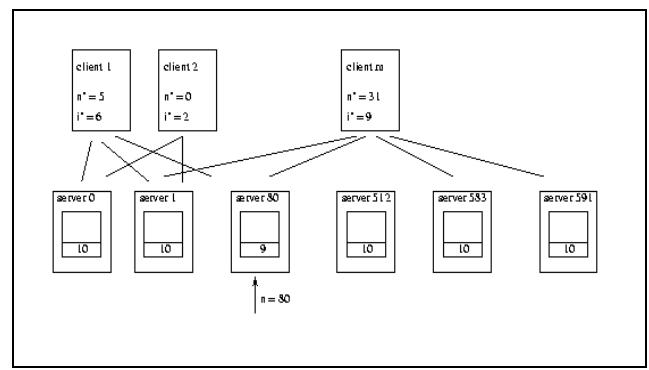

Fig. 1 : LH* principe

Every client maintains his picture that consists of the two indices i' and n'; where i' is the level of his hashing function, n' is the pointer of the next server that must split. The client sends his request (insertion, modification, suppression either or update) relative to the key c to the server m determined by the algorithm A1 (the address is not necessarily correct).

$m \leftarrow hi'(c) = c \bmod 2^{i'}$
if $m < n'$
$m \leftarrow hi'+1\ (c) = c \bmod 2^{i'+1}$

*Algorithm A1*

The server m that receives the client's request, must apply his hashing function. If the result is different of the server's number in question, the request is redirected to another server. (Algorithm A2).

$a' \leftarrow hj(C)$
if $a' \neq m$
$\quad a'' \leftarrow hj\text{-}1(C)$
$\quad$ if $a'' > a$ et $a'' < a'$
$\quad\quad a' \leftarrow a''$

*Algorithm A2*

In case of redirection, an adjustment message will be addressed to the client so he brings his picture update (Algorithm 3)

1: if $i > i'$
$\quad i' \leftarrow i\text{-}1$
$\quad n' \leftarrow a+1$
2: if $n' \geq 2i'$
$\quad n' \leftarrow 0$
$\quad i' \leftarrow i'+1$

*Algorithm A3*

The LH* file increase by the linear manner, to every collision, a message is sent to the split coordinator which sends the split order to the n server.
For more details of LH* algorithms, the reader can refer to [10] [11].

## 3. Trie Hashing (TH)

Trie hashing [9] is one of the fastest access methods for dynamic and ordered files. Its efficiency lies in the use of a trie(Fig 2),. It starts out with a bucket in which all keys will be stored. When an overow occurs, another bucket will be appended at the end of the primary file. All keys will then be redistributed into the overow bucket and the new bucket just allocated by comparing the value of the first character of each key with a discriminator which is a suitable value that will usually divide the keys evenly. A key having the first character smaller than or equal to the discriminator will go into the original bucket, otherwise it will go into the new bucket No secondary file is needed.

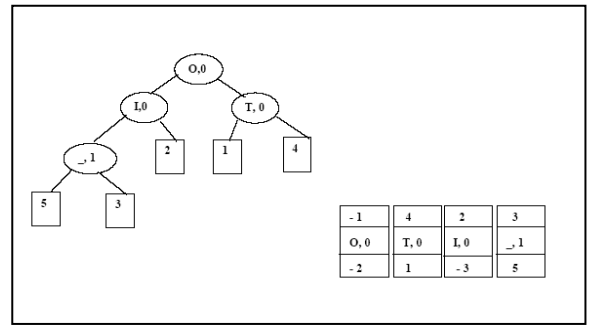

Fig 2 : Trie Hashing Principe

The result of splitting the buckets is described in a trie with the discriminator and its associated position within the key stored in each internal node, and the bucket addresses stored in the leaf nodes. When the keys are numbers, a bit is used for comparison instead of using the whole character. As a result, the discriminators are not required to be stored in the internal nodes. During the search, each bit of the given key will be examined. If it is zero, proceed to the left subtree otherwise go to the right subtree.This is the digital searching .

It is possible that after redistribution, all keys go into the same bucket and overow again. This may result in multiple empty buckets being allocated and the depth of the trie will be increased by more than one. If the keys are uniformly distributed, these empty buckets will be filled subsequently.

We may describe the bit checking by a family of functions {sd} where sd(k) = (k=2d) mod 2, d is the depth of the node in which sd is being used. Below are the algorithms used to searcher and insert a key k.

```
Searcher (k)
Trienode p
p← the root of the trie
d←0
while (p I an internal node )
            if  s_d(k) =0
                    p ← p.left
                else
                       p← p.right
                d ← d+1
return (p)
```

Searcher algorithm

```
Insert(k)
p←searcher(k)
Read in p:bucket
 If (p:bucket is not full)
          insert k
    else
        Allocate one more bucket
  Perform bucket splitting and update the trie
```

Insertion algorithm

## 4. Principle of LH*TH

LH*TH is a variant of LH * using two levels of indexing. The 1st is network index managed by algorithms A1, A2, A3 of the LH* diagram that permit to the client to find the server (LH* bucket) containing the desired information. The 2nd is local index managed by the TH's algorithms [9]. Every LH* bucket (server) contains several THs buckets

4.1 Internal organization of an LH*TH bucket:

A LH*TH bucket is composed of the trie (hashing function) and the TH buckets. Initially only one bucket exists; that is the TH0 bucket, and the trie is composed of only one node indexing TH0 bucket. From a b capacity (definite in advance), the bucket TH0 splits and a bucket TH1 is created, and so forth. This splitting is internal to every LH* server. It bases on the local trie, which is brought update after every internal splitting. The Figure (Fig. 3) shows the internal architecture of an LH*TH bucket.

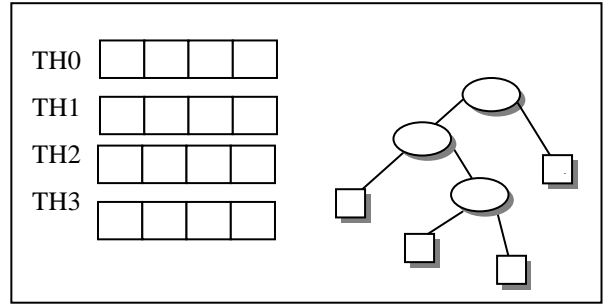

Fig 3: internal architecture interne of an LH*TH bucket.

### 4.2 Evolution of a SDDS LH*TH file

Either inserted the following key sequence: 320, 11, 10, 25, 31, 54, 126, 219, 250, 251, 280, 13, 322, 120, Under the following hypothesis: b = 4 (capacity of the TH bucket) and k=4 (number of TH bucket slot in the LH* server) The insertion of keys: 320, 11, 10, 25 are placed in TH0 bucket; the insertion of the key 31 provokes a collision on the TH0 bucket that will split while using the algorithm to TH1 bucket. And so forth until the key 322. The figure (Fig 4) shows the content of the LH*TH buckets without the key 120.

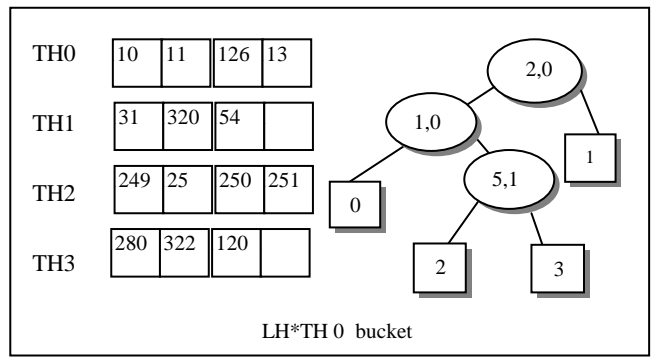

Fig 4 : the LH*TH file state before splitting

The insertion of the last key 120 provokes a general collision on LH*TH bucket, the split will be treated by the algorithm of the LH *. The figure Fig 5 gives the state of the LH*TH file after splitting:

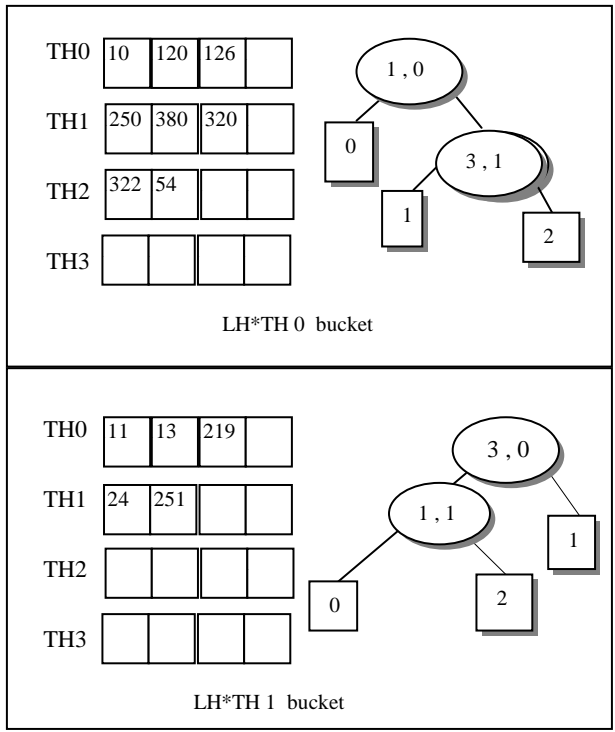

Fig 5: the LH*TH file state after splitting

## 5. PERFORMANCES

This section presents the performances study of the new SDDS LH*TH. we begin by presenting profile of the system on which we have achieved tests, then we will present results of the parameter observed that are load factor and different coast of the access operations.

### 5.1 Experimental environment:

We have implemented the SDDSs LH*TH and LH * on multi computer composed of 4 PC executing the LINUX system (Mandrake 81) and connected as local network by a Switcher 100 Mb/ses. Every machine can be client and/or server.
The experiment that we achieved shows that performances of access are not influenced by the size of file bucket. The following exposed results are obtained with size of 1000 articles by bucket.

### 5.2 Load factor:

The table Tab1 summarizes the tests on the load factor. We can conclude that the load factor of the LH*TH file varies between 55% and 75%, it is practically same as the one of LH *.

| Article number | Load factor | |
|---|---|---|
| | LH* | LH*TH |
| 10000 | 0,734 | 0,667 |
| 20000 | 0,605 | 0,510 |
| 30000 | 0,656 | 0,630 |
| 40000 | 0,595 | 0,597 |
| 50000 | 0,671 | 0,559 |
| 60000 | 0,746 | 0,657 |
| 70000 | 0,804 | 0,758 |
| 80000 | 0,493 | 0,448 |
| 90000 | 0,507 | 0,506 |
| 100000 | 0,680 | 0,666 |
| Average | 0.649 | 0.601 |

Tab 1: load factor

### 5.3 Insertion:

To value the average time of an insertion we have launched the creation operation of the LH*TH (resp.LH *) with number of different article (10000,20000…100000), and at every insertion the time of answer is valued. The table Tab 2 illustrate the average insertion time according to the number of inserted articles. It is the order of 0,85 Ms/insertion for the LH* file and 0,87 Ms/insertions for the one of LH*TH. One may notice that the average time of an insertion for LH*TH is more important than the one of LH*.this is due to the digital tree maintenance. While increasing the size of the file the time of insertions remains practically steady: insertions are scalable.

| Articles number | insertion Average time (Ms) | |
|---|---|---|
| | LH* | LH*TH |
| 10000 | 0,722 | 0,767 |
| 20000 | 0,860 | 0,901 |
| 30000 | 0,894 | 0,913 |
| 40000 | 0,871 | 0,904 |
| 50000 | 0,882 | 0,944 |
| 60000 | 0,845 | 0,850 |
| 70000 | 0,856 | 0,859 |
| 80000 | 0,870 | 0,875 |
| 90000 | 0,876 | 0,885 |
| 100000 | 0,886 | 0,892 |
| Average | 0,856 | 0,879 |

Tab 2: the insertion average time

5.4 The number message of an insertion:

To observe the behaviour of number of message exchanged in the Multi-Computer, we have achieved an experience under the same conditions as in section 5.3. Table tab3 shows the average number of messages per insert. It is practically the even for the two SDDSs

| Articles number | The average message number of the insertion | |
|---|---|---|
| | LH* | LH*TH |
| 10000 | 2,36 | 2,50 |
| 20000 | 2,40 | 2,39 |
| 30000 | 2,34 | 2,34 |
| 40000 | 2,32 | 2,35 |
| 50000 | 2,31 | 2,37 |
| 60000 | 2,32 | 2,39 |
| 70000 | 2,32 | 2,40 |
| 80000 | 2,31 | 2,37 |
| 90000 | 2,31 | 2,38 |
| 100000 | 2,31 | 2,40 |
| Average | 2,330 | 1,389 |

Tab 3: The insertion average message number

3.1 Times of a research:

In the table Tab 4 we present results of tests of the average research time of an article in the multi-computer. The conditions of the tests were as follows: a client r launches a set of 10000 article research then 20000 until 100000 and to every research we calculate the answer time. From obtained results, we can conclude that the SDDS LH*TH gives very interesting research with regard to the one of LH *. Note that the SDDS LH*TH permits a gain of time of 0.3 MSS by research operation.

| Articles number | Searcher average time | |
|---|---|---|
| | LH* | LH*TH |
| 10000 | 0,50496 | 0,30634 |
| 20000 | 0,45466 | 0,30151 |
| 30000 | 0,47163 | 0,28572 |
| 40000 | 0,65496 | 0,28268 |
| 50000 | 0,58557 | 0,26345 |
| 60000 | 0,52163 | 0,26575 |
| 70000 | 0,49067 | 0,27439 |
| 80000 | 0,44965 | 0,27307 |
| 90000 | 0,39969 | 0,27602 |
| 100000 | 0,46301 | 0,25963 |
| Average | 0,49964 | 0,27886 |

Tab 4 : average time of key search

# 6. Conclusions

Nowadays, the technology of multi computer is among the most promising research topics in data processing whose repercussions will be fundamental. It is notably about the technology of specific and more effective data structures as the SDDSs.

In this paper we presented the new SDDS LH*TH that is based on LH * as external hashing function and the Trie hashing as internal function.

Our implementation has been achieved on a multi-computer functioning with system Linux (Mandrake 8.1)
the measures of presented performances demonstrated that in LH*TH an insertion is achieved with one time of access of the order 0.87 Mses, a research is done in 0.27 MSS and load factor is upper that 65%. Noting that all the operations on the SDDS LH*TH are scalables.

The comparative analysis between LH*TH and LH * has show that the new SDDS LH*TH preserves all properties of LH * with the advantage of research that is distinctly Better in LH*TH.

The future works must interested, on one hand, In the parallel and intervals request, on the other in the integration of the new SDDS in SGF and SGBD distributed. Finally, we note that a SQL-LH*TH version is in progress of implementation

**Acknowledgments**

The author would like to thank Pr Zegour Djamel Eddine Director of LCSI and Pr Litwin Witold Director of CERIA for taking the time discusses the ideas presented here.

**Dr ARIDJ Mohamed** is an Associate Professor at the Computer Science Department of Chlef University Algeria. His current research area includes distributed systems, multi computers, Distributed hashing, Scalable Distributed Data Structures (SDDSs), access method .He received his Doctorate degree in distributed systems applications from Ecole Supérieur d'Informatique (ESI) Oued Smar Algeria.